# Exceptional Point of Sixth Order Degeneracy in a Modified Coupled Resonators Optical Waveguide

Mohamed Y. Nada and Filippo Capolino

*Department of Electrical Engineering and Computer Science, University of California, Irvine, CA 92697, USA*

We demonstrate for the first time the occurrence of a *sixth* order exceptional point of degeneracy (EPD) in a realistic multimode optical photonic structure by using a modified periodic coupled resonators optical waveguide (CROW), at the optical wavelength $\lambda_e = 1550\,\mathrm{nm}$. The 6th order EPD is obtained in a CROW without the need of loss or gain and such an EPD corresponds to a very special band edge of the periodic photonic structure where six eigenmode coalesce, so we refer to it as the 6th order degenerate band edge (6DBE). Moreover, we report a new scaling law of the quality factor $Q$ of an optical cavity made of such a periodic 6DBE-CROW with cavity length as $Q \propto N^7$, when operating near the 6DBE with $N$ being the number of unit cells in the periodic finite-length CROW. Furthermore, we elaborate on the application of the 6DBE to ultra-low-threshold lasers. We present a novel scaling law of the lasing threshold that scales as $N^{-7}$ when operating near the 6DBE. Also, we show the superiority of the threshold scaling of the 6DBE-CROW to the scaling of another CROW with the same size operating near a 4th order EPD that is often referred to as the degenerate band edge (DBE). The lasing threshold scaling of the DBE-CROW laser is shown here for the first time. We also discuss the high sensitivity of the proposed 6DBE-CROW to perturbations, that may find applications in sensors, modulators, optical switches, nonlinear devices, and $Q$-switching cavities.

## I.    Introduction

An exceptional point of degeneracy (EPD) is a point at which two or more system eigenmodes coalesce in both eigenvalues and eigenvectors [1]–[5]. Since the characterizing feature of an exceptional point is the strong degeneracy of at least two eigenmodes, as implied in [6], we stress the importance of referring to it as "degeneracy" and including the "D" in EPD. Despite most of the published work on EPDs is related to PT symmetry [3], [4], the occurrence of an EPD actually does not require a system to satisfy PT symmetry. Indeed, EPDs have been recently found also in single resonators by just adopting time variation of one of its components [7]. In this paper we focus on an EPD of sixth order in a *lossless and gainless photonic structure*, where the order of the EPD is determined by the number of coalescing eigenmodes at the exceptional point. EPDs in lossless structures [8]–[13] are associated with slow-wave phenomena, including band edges, whereby the group velocity of the propagating wave is almost vanishing [14]–[16]. In the optical realm, slow-wave phenomenon has invaded many intriguing aspects of optical resonators in which nonlinearities [17], and gain/absorption [18], among other features, can be significantly enhanced. Generally, the existence of an EPD in electromagnetic systems leads to unique properties that cannot be obtained in conventional structures such as the giant field enhancement, strong enhancement in the local density of states [12], unconventional scaling of the quality factor (Q factor) with structure length [11]–[13], [19], and extreme sensitivity to perturbations [7], [20]. Such properties can be utilized in various applications like modulators, switches, high quality factor resonators and sensors.

Optical sensors based on the confinement of light in optical microresonators (or microcavities) have received a surge of interest nowadays [21], [22]. Sensors based on slow light in optical microcavities [23], [24] require the usage of high Q factor resonators [25], [26], which can be done by designing the optical resonators to operate near an EPD [13]. High Q resonators are also beneficial for other different applications including filters [27], optical switching [28], optical delay line devices [29], and lasers [30]. On the other hand, cascading a chain of coupled micro resonators, as shown in [31], has stimulated a great interest in studying coupled resonator optical waveguides (CROWs) as special devices for slow light transport [29], [5].

We stress that the EPD presented in this paper is obtained in periodic structures without the existence of loss or gain in the system. The simplest EPD is the one that exists at the band edge of any periodic structure due to the coalescing of two eigenmodes. Such 2nd order EPD is referred to as the regular band edge (RBE) [8]. The 3rd order EPD was found in non-reciprocal structures [2], [32] and it is often referred to as the stationary inflection point (SIP). Recently the SIP has been shown in lossless, reciprocal structures such as a 3-way waveguide [33] and the modified CROW [5]. The 4th order EPD is a band edge of a periodic structure, so we refer to it as the degenerate band edge (DBE). Such an EPD was explored in various structures [8], [11], [12], [34] and it was even found in CROWs [5], [13]. Higher orders EPDs were studied theoretically in [10] assuming coupled mode theory without referring to any particular structure, and to the best of our knowledge, EPDs of high orders (greater than 4) had not been found in any realistic optical structure yet.

In this paper, we present a realistic optical structure made of a CROW side coupled to a waveguide that is capable of exhibiting a 6th order EPD in its dispersion diagram. Throughout the paper we refer to this 6th order EPD as 6DBE since it is obtained at the band edge of the CROW without the need of loss or gain. We rely on the general formulation provided in [5], and the several contributions presented here are summarized as follows. We provide a design with the dimension of the coupled ring resonators, the effective refractive indices, and practical values of coupling coefficients for the CROW to exhibit a 6DBE in its dispersion diagram. We also show some of the unique properties of CROWs operating near the 6DBE such as the scaling of the quality factor with cavity length. Indeed, we report a new quality factor scaling law as $Q \propto N^7$ with $N$ being the number of unit cells in the periodic finite-length cavity. Finally, we explore two possible interesting applications of our modified CROW structure operating in close proximity to the 6DBE: sensors and low threshold lasing. In the latter application we show a newly observed scaling law of the lasing threshold that decreases as $N^{-7}$ when operating near the 6DBE. Moreover, we compare the threshold scaling of a finite-length CROW cavity operating near the 6DBE to the scaling of another CROW-cavity, of the same dimensions, operating near the 4DBE. In doing so, we also describe for the first time the lasing threshold of a cavity operating near a 4DBE based on this CROW geometry.

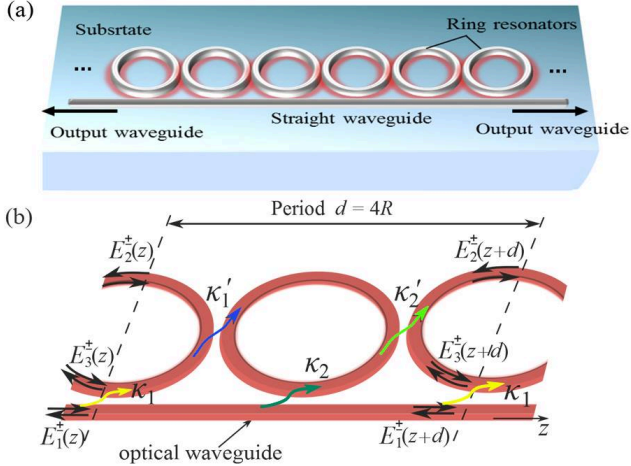

FIG. 1. (a) The 6DBE-CROW consists of a chain of coupled ring resonators of radius $R$ that is side coupled to a rectangular straight waveguide. The 6DBE-CROW is periodic in the $z$-direction with a period $d = 4R$. (b) A unit cell of the periodic 6DBE-CROW with the electric field wave amplitudes defined at the periodic-cell boundaries. The field coupling coefficients between the coupled rings are alternating between $\kappa_1'$ and $\kappa_2'$ whereas the field coupling coefficients between the straight optical waveguide and the rings are alternating between $\kappa_1$ and $\kappa_2$.

## II.  CROW with 6th Order EPD

We introduce the modified coupled resonator optical waveguide (CROW) structure shown in Fig. 1(a) which is capable of exhibiting various orders of EPDs. This structure represents a three-way waveguide, i.e., there are three propagating or evanescent modes in each positive and negative $z$-direction. The theory of EPDs in CROWs was discussed in [5], where we have discussed degeneracies of orders 2,3 and 4. In this paper we show the first realization of a waveguide that exhibits a 6th order EPD in its dispersion diagram, which is the maximum degeneracy order with a three-way waveguide. The structure is made of a chain of coupled ring resonator optical waveguides where the field coupling coefficients are alternating from one ring to another as $\kappa_1'$ and $\kappa_2'$, and the outer radius of each ring is $R$. The CROW is side-coupled to a uniform optical waveguide with alternating field coupling coefficients $\kappa_1$ and $\kappa_2$ as shown in Fig. 1. We call this modified CROW, designed to exhibit the 6th order EPD, as 6DBE-CROW.

Following [5], we assume a single mode propagation in each waveguide segment, and the eigenwaves' phase propagation in the positive/negative $z$-direction of the waveguides and the rings is represented by $e^{\pm i n_w k_0 z}$ and $e^{\pm i n_r k_0 z}$, respectively. Here, $n_w$, $n_r$ are the effective refractive indices of the waveguide and the rings and $k_0 = \omega / c$ is the wavenumber in free space. Also, throughout the paper we assume that the time convention is $e^{-i\omega t}$. Each ring resonator external radius is $R$ so that the 6DBE-CROW is periodic with a period $d = 4R$ (the unit cell consists of 2 rings), where for simplicity we neglect the gap dimensions between adjacent rings as was done in [5], [13], [14].

To explore the unique modal characteristics of this 6DBE-CROW, we represent the wave propagation along $z$ using complex electric field wave amplitudes that are defined as shown in Fig. 1(b), at the immediate left of cell boundaries, i.e., just before where the ideal coupling $\kappa_1$ occurs. Therefore, at any point $z$, there are three complex electric field wave amplitudes that propagate in the positive $z$-direction described by the 3-dimensional vector $\mathbf{E}^+(z) = \begin{bmatrix} E_1^+(z), & E_2^+(z), & E_3^+(z) \end{bmatrix}^T$ and another three field wave amplitudes that propagate in the negative $z$-direction represented by the vector $\mathbf{E}^-(z) = \begin{bmatrix} E_1^-(z), & E_2^-(z), & E_3^-(z) \end{bmatrix}^T$. We define a state vector composed of the six field amplitude components as

$$\boldsymbol{\psi}(z) = \begin{bmatrix} \mathbf{E}^+(z) \\ \mathbf{E}^-(z) \end{bmatrix} \qquad (1)$$

Using coupled mode theory [5], [14], [35] and assuming single mode propagation in each segment (for each direction), the evolution from cell to cell of the state vector is governed by the equation $\boldsymbol{\psi}(z+d) = \underline{\mathbf{T}} \, \boldsymbol{\psi}(z)$, where d is the period, and $\underline{\mathbf{T}}$ is a $6 \times 6$ transfer matrix representing the evolution across a unit cell. The expression of the unit cell transfer matrix $\underline{\mathbf{T}}$ and a related discussion is in Appendix B. We utilize this transfer matrix formalism to investigate the evolution of the state vector along the 6DBE-CROW hence we derive the eigenwave characteristics. We obtain the $\omega - k$ dispersion relation of the 6DBE-CROW eigenmodes as

$$D(k,\omega) = \det[\underline{\mathbf{T}} - \zeta \underline{\mathbf{1}}] = 0; \quad D(k,\omega) = \sum_{l=0}^{6} c_l(\omega)\zeta^l \quad (2)$$

where $\zeta = \exp(ikd)$, $k$ is the Bloch wavenumber along $z$, $\omega$ is the angular frequency, and $\underline{\mathbf{1}}$ is the $6 \times 6$ identity matrix. The polynomial coefficients $c_l(\omega)$ are function of frequency and the 6DBE-CROW parameters. It is clear from the dispersion relation (2) that at any frequency, there exist six Bloch eigenmodes guided by the 6DBE-CROW and we showed in [5] that there are particular frequencies at which some eigenmodes coalesce in both their wavenumber and eigenvectors. The number of coalescing eigenwaves at the EPD represents the order of the EPD. In [5] we have shown different designs where the proposed CROW can support 2nd, 3rd and 4th order EPDs. However, in this paper we show that such CROW can exhibit also a 6th order EPD in its dispersion diagram, which is the highest order that can be obtained using such structure. When such a condition occurs, the six eigenvectors of $\mathbf{T}$ coalesce as discussed in the next section. It is important to point out that this is the first time when the 6th order EPD is shown in an optical realistic structure.

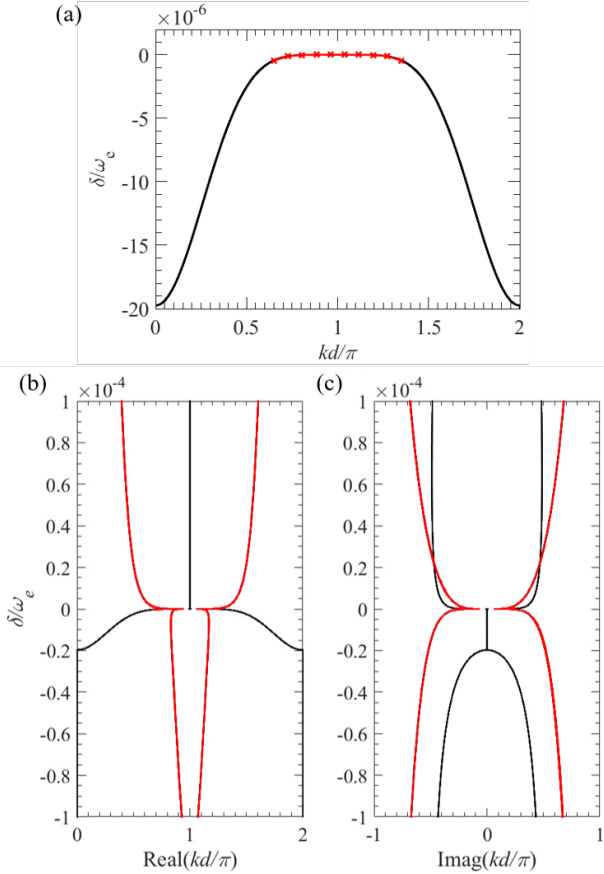

FIG. 2. Floquet-Bloch wavenumber dispersion diagram of the 6DBE-CROW whose unit cell is shown in Fig.1(b). (a) Only the propagating modes with purely real wavenumber values are shown, where the 6DBE occurs at the frequency corresponding to the wavelength $\lambda_e = 1550$ nm and $\delta = \omega - \omega_e$. The red cross markers show the dispersion fitting formula given by equation (5) with the fitting flatness parameter $\eta_e = 0.00025$. (b) and (c) Complex dispersion diagram showing both real (b) and imaginary (c) parts of the six Floquet-Bloch wavenumbers $k$ versus normalized real frequency $\delta / \omega_e$. This diagram shows six wavenumbers coalescing at $\omega = \omega_e$. Branches with two overlapping modes are in red color. The parameters of the unit cell are chosen as the radius is $R = 13 \,\mu$m, the cross coupling coefficients are $\kappa_1 = 0.89$, $\kappa_1' = 0.03$, $\kappa_2 = 0.4$, $\kappa_2' = 0.72$, and the effective refractive indices are $n_r = 2.45$, and $n_w = 2.5$.

## III. Mathematical Description of The Sixth Order EPD

When an electromagnetic system exhibits a sixth order EPD in its dispersion diagram, exactly at the EPD frequency the unit-cell transfer matrix $\underline{\mathbf{T}}_e$ contains six degenerate eigenvalues $\zeta_e$, where the $e$ subscript denotes the EPD. There are six degenerate eigenvectors, therefore the algebraic multiplicity of $\zeta_e$ is 6 but its geometrical multiplicity is 1. Hence, the T-matrix $\underline{\mathbf{T}}_e$ is similar to a Jordan matrix of order 6 which is represented as

$$\underline{\mathbf{T}}_e = \underline{\mathbf{V}} \underline{\boldsymbol{\Lambda}}_e \underline{\mathbf{V}}^{-1}, \quad \underline{\boldsymbol{\Lambda}}_e = \begin{pmatrix} \zeta_e & 1 & 0 & 0 & 0 & 0 \\ 0 & \zeta_e & 1 & 0 & 0 & 0 \\ 0 & 0 & \zeta_e & 1 & 0 & 0 \\ 0 & 0 & 0 & \zeta_e & 1 & 0 \\ 0 & 0 & 0 & 0 & \zeta_e & 1 \\ 0 & 0 & 0 & 0 & 0 & \zeta_e \end{pmatrix} (3)$$

Here $\underline{\mathbf{V}}$ is the similarity transformation matrix whose columns comprise one regular eigenvector and five generalized eigenvectors corresponding to the six degenerate eigenvalue solutions $\zeta_e = \exp(ik_e d)$ where $k_e$ is the wavenumber at the 6DBE and $\underline{\boldsymbol{\Lambda}}_e$ is a 6×6 Jordan block.

Due to reciprocity, the six Bloch wavenumber solutions of (2) have to form pairs made of negative and positive values, i.e., $k$ and $-k$ are both solutions. Since there are only six solutions, the 6DBE has to be at $k_e$ and $-k_e$ simultaneously, i.e., $k_e = -k_e$, which leads to only two possibilities for the $k_e$ value: either $k_e = 0$ or $k_e = \pi / d$, i.e., either at the edge or center of the first Brillouin zone (BZ), respectively (defined here by the interval $\begin{bmatrix} 0 & 2\pi / d \end{bmatrix}$). In other words, this sixth order EPD cannot occur at any other point of the BZ in this reciprocal system. This follows that the dispersion relation (2), when fixing the frequency exactly at the 6DBE one, must be in the form

$$(\zeta_e \pm 1)^6 = 0 \qquad (4)$$

where the + sign corresponds to the case when $k_e = 0$ and the − sign corresponds to the case when $k_e = \pi / d$; which is the case discussed next.

There are many possible sets of parameters in to realize the 6DBE that can be obtained by proper tuning of the coupling coefficients, effective refractive indices and radius of the rings. To facilitate the tuning of the design parameters, we equate the coefficients of the polynomial (2) at the desired 6DBE frequency to those of (4). Hence, we can find a set of necessary equations governing the choice of the different 6DBE-CROW parameters that can be solved numerically to determine the different parameters. Note that equating the coefficients of the polynomials (2) and (4) at a specific desired optical frequency provides a necessary condition to find the 6DBE at that desired frequency, but the sufficient condition for the 6DBE to exist is to check that the system six eigenvectors are coalescing at such desired 6DBE frequency.

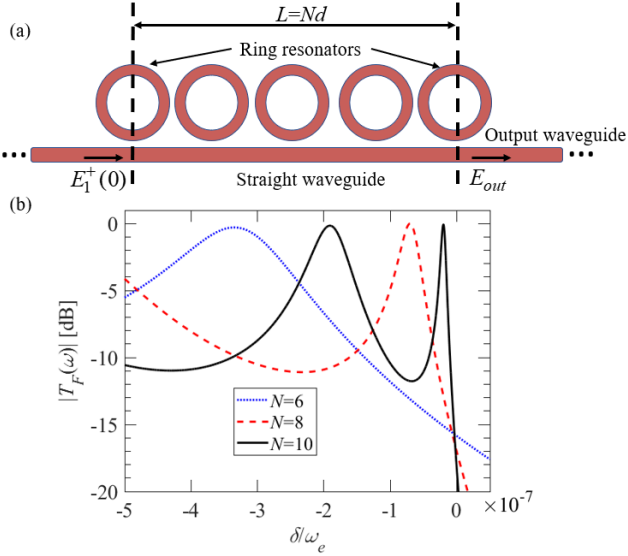

Fig. 3. (a) Cavity made of a finite-length 6DBE-CROW of length $L=Nd$, where $N$ is the number of unit cells. The straight waveguide is extended on both sides working as input and output waveguide without any discontinuity. The 6DBE-CROW is excited from the left extended waveguide by the field amplitude $E_1^+(0)$ and the output electric field amplitude exiting the waveguide from the right is $E_{out}$. (b) Magnitude of the transfer function, $|T_F(\omega)|$, in dB of a 6DBE-CROW cavity operating in close proximity of the 6DBE frequency calculated for three different number of unit cells ($N$) of the 6DBE-CROW given as 6, 8 and 10. Resonances of such a cavity, denoted by transmission peaks, gets sharper when they get closer to the 6DBE frequency, i.e., when $\delta = \omega - \omega_e \to 0$. The longer the cavity, the closer the resonances to the 6DBE frequency.

In all the subsequent analysis and results, the 6DBE wavelength is designed to be at the optical wavelength $\lambda_e = 2\pi c / \omega_e = 1550$ nm where $\omega_e$ is the 6DBE angular frequency. We have solved the necessary equations and found the 6DBE-CROW parameters that may lead to the existence of the 6DBE and we have confirmed the existence of the 6DBE by the coalescence of the six eigenvectors. The parameters of the unit cell are as follows: radius is $R = 13\,\mu\text{m}$, the cross coupling coefficients are $\kappa_1 = 0.89$, $\kappa_1' = 0.03, \kappa_2 = 0.4, \kappa_2' = 0.72$, and the effective refractive indices are $n_r = 2.45$, and $n_w = 2.5$. The used coupling parameters and refractive indices values are practical, and they lie within the range of values that have been realized using silicon on insulator (SOI) technology [36]–[38]. The dispersion diagram of this unit cell is shown in Fig. 2, where in Fig. 2(a) we show only the propagating modes, i.e., modes with zero imaginary part of the Bloch wavenumber $k$, while in Figs.

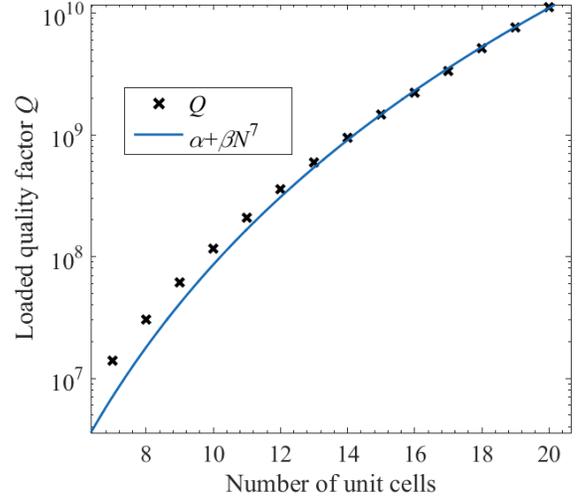

Fig. 4. Loaded quality factor ($Q$) of the lossless 6DBE-CROW calculated at different cavity lengths, i.e., for growing number of unit cells $N$. The values of $Q$, denoted by black cross symbols, are calculated using the group delay method while the blue solid line represents the fitting curve with equation $Q = \alpha + \beta N^7$.

2(b) and (c) we show the complete complex dispersion diagram in order to see clearly the coalescence of the six modes at the EPD. Branches with two overlapping modes are shown in red. In the dispersion diagram in Fig. 2, the 6DBE point is characterized by $\omega_e$ and $k_e = \pi / d$, and in its vicinity the dispersion is well approximated by

$$(1 - \omega / \omega_e) \approx \eta_e (1 - k / k_e)^6 \qquad (5)$$

where $k_e = \pi / d$ is the 6DBE wavenumber at the band edge. In analogy to the theory presented in [13], the dimensionless "flatness parameter" $\eta_e$ is related to the value of the sixth derivative of $\omega$ with respect to $k$ at the 6DBE angular frequency $\omega_e$, i.e.,

$$d^6 \omega / dk^6 = -\eta_e \omega_e / \left( 720 k_e^6 \right) \qquad (6)$$

and it dictates the flatness of the dispersion relation at $\omega_e$. Such fitting formula is shown with red cross markers in Fig. 2(a) with the flatness parameter $\eta_e = 0.00025$. The flatness parameter plays an important role in the scaling of the quality factor of the whole structure as explained in [13]. The dispersion diagram in Fig. 2 shows only the coalescence of the six eigenvalues but not the coalescence of the eigenvectors. To quantify the coalescence of the eigenvectors and define how close the system is to an EPD, we use the *hyperdistance* concept [20] shown in Appendix C.

## IV.    Giant Resonance in CROW Cavity with Sixth Order EPD

We start this section by showing the transfer function of a cavity made of a lossless finite-length 6DBE-CROW comprises of $N$

unit-cells, shown in Fig. 3(a), as a function of the angular frequency $\omega$ for different numbers of unit cells. We recall that a unit cell is made of two rings, since the coupling coefficients are alternating from one ring to its adjacent one, as shown in Fig. 1. A unit cell starts at the center of the ring just before the coupling point with the straight waveguide, as shown with a dashed line in Fig. 3(a). At each end, at $z = 0$ and $z = L$, there are three ports to be terminated: the first and last rings are terminated with half rings; whereas the straight waveguide is extended, without changing the waveguide dimensions, for both $z < 0$ and $z > L$ to define the input and the output of the CROW. The transfer function is defined as

$$T_F(\omega) = \frac{E_{out}}{E_1^+(0)} \qquad (7)$$

where $E_{out}$ is the electric field wave amplitude escaping the straight output waveguide from the right at $z = L$, while $E_1^+(0)$ is the field amplitude of the incident-wave to the waveguide from left. To calculate $T_F(\omega)$, we first obtain the state vector at the right boundary of the last unit cell, just before the coupling point between the waveguide and ring resonators, as $\Psi(z = L) = \underline{\mathbf{T}}^N \Psi_0$. Here $\underline{\mathbf{T}}$ is the T-matrix of one unit-cell and $\Psi_0$ is the state vector defined just before the coupling point of the left boundary of the first unit cell. Then we apply the boundary conditions at both ends of the finite length CROW as in Eq. (30) and Eq. (31) in [5] to get $\Psi(z = L)$, hence we calculate $E_{out} = \sqrt{1 - \kappa_1^2} E_1^+(L) + i\kappa_1 E_3^+(L)$. Note that $E_{out}$ is defined on the right side of the coupling point, at $z = L$ as shown in Fig. 3. The magnitude of the transfer function $T_F(\omega)$ is shown in Fig. 3(b), and it is clear that the cavity has multiple resonances, where resonances are denoted by transfer function peaks. We identify the resonance frequency closest to the 6DBE frequency (the closest transmission peak) as the 6DBE *resonance*, denoted by $\omega_{r,e}$. Such a resonance exhibits the narrowest spectral width (highest $Q$ factor), as seen for the case of $N = 10$ unit cells in Fig. 3(b). Also, one can see that the 6DBE resonance gets sharper (higher $Q$ factor) and it gets closer to $\omega_e$ by increasing the number of unit cells $N$ of a cavity. The 6DBE resonance frequency follows the asymptotic formula

$$\omega_{r,e} / \omega_e \simeq 1 - \eta_e / N^6 \qquad (8)$$

in analogy to what was discussed in [8], [13] for a fourth order degeneracy.

Now, we analyze the scaling of the $Q$ factor with the length of a 6DBE-CROW cavity where the straight waveguide continues without terminations or discontinuities on both left and right ends of the waveguide, as shown in Fig. 3(a). The loaded $Q$ factor of the cavity versus the number of unit cells $N$ is shown in Fig. 4, where the $Q$ factor is calculated by $Q = \omega_r \tau_g / 2$ in which $\tau_g$ is the group delay defined as the derivative of the transfer function phase, $\angle T_F$, with respect to the angular frequency $\omega$, i.e.,

$\tau_g = \partial \angle T_F / \partial \omega$ [5], [12], [13]. The scaling of the $Q$ factor versus $N$ is fitted by $\alpha + \beta N^7$ where the fitting parameters $\alpha$ and $\beta$ for the case shown in Fig. 4 are given as $\alpha \approx 27.3$ and $\beta \approx 8.5$. Hence, we show here that the 6DBE-CROW cavity made of a waveguide structure of finite length has a $Q$ factor that asymptotically grows with length as

$$Q \simeq \beta N^7. \qquad (9)$$

The theoretical justification of the $N^7$ scaling can be understood from the expression of the $Q$ factor where, as stated earlier, $Q$ is directly proportional to the group delay $\tau_g = L / v_g$, with $L = Nd$ being the length of the finite length cavity and $v_g$ being the group velocity. The group velocity is proven to be inversely proportional to $N^m$, where $m$ is the order of the EPD [12], [39]. Hence, the group delay and in turn the Q factor is proportional to $N^{m+1}$, where $m = 6$ for the 6DBE.

## V. Applications of 6DBE-CROWs Operating Near a 6DBE

We show two possible applications associated with the finite-length 6DBE-CROW operating near the 6DBE: an ultra-low-threshold laser and an ultra-sensitive sensor.

### A. Low threshold optical oscillator

In this section we explore an interesting application of the 6DBE-CROW which is a low threshold lasing oscillator. To investigate the lasing threshold in a finite 6DBE-CROW, we introduce distributed gain only in the ring resonators of a finite lossless 6DBE-CROW. The distributed gain may be introduced by doping the rings or their surrounding with optically pumped active atoms, e.g., $Er^{3+}$, or by using layers of quantum wells [40]. Here we model the distributed gain as a negative imaginary part of the effective refractive index of the ring resonators, i.e., $n_r = n_{real} + in_{imag}$, where in the introduced design $n_{real} = 2.45$ and $n_{imag}$ is dictated by the concentration of the active material and the pumping power. Hence, we define the per unit length propagation power gain as $\gamma = -2k_0 n_{imag}$ [m$^{-1}$], where $k_0$ is the free space wavenumber [12] (also see Ch.5, P. 229 in [41]). Note that if we introduce large gain into the 6DBE-CROW, this would deteriorate the degeneracy and the unique properties associated with the slow-wave phenomena, as discussed in [40], [42]. So, we start with small values of distributed gain and increase it gradually, monitoring when the structure starts oscillating. This is done by tracking the complex-frequency poles loci of the transfer function $T_F(\omega)$ [43] when varying the distributed propagation gain values, and the method is summarized in Appendix A. Hence, we calculate the lasing gain threshold $\gamma_{th}$ defined as the minimum amount of distributed propagation gain (assumed uniform along the structure) that is sufficient to maintain lasing in the 6DBE-CROW cavity through stimulated emission. In Fig. 5 we show the scaling of the lasing threshold versus number of unit cells $N$ constituting the finite-length 6DBE-CROW, which is shown with black markers. The extraordinary property of the 6DBE-CROW is that the lasing threshold decreases with

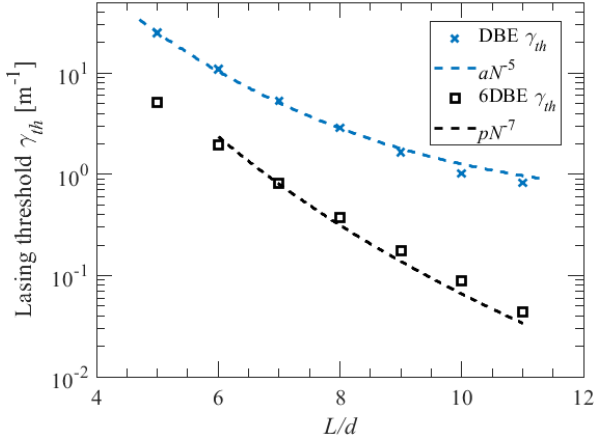

Fig. 5. Lasing threshold of a finite-length 6DBE-CROW calculated for different lengths of the structure. Blue markers represent the calculated threshold for a cavity operating near the DBE (4th order EPD) whereas black markers are the calculated threshold for a cavity operating near the 6DBE. Dashed lines represent the fitting of these data: the blue one represents the scaling $\gamma_{th,\text{DBE}} \simeq aN^{-5}$ of the DBE laser [32], whereas the black one represents the asymptotic scaling $\gamma_{th,6DBE} \simeq pN^{-7}$ of a 6DBE laser.

increasing number of unit cells $N$, i.e., increasing the structure length, following the unique trend

$$\gamma_{th} \simeq pN^{-7} \qquad (10)$$

where $p$ is a fitting constant, as shown with the dashed black fitting curve in Fig. 5. This result is quite expected as the scaling of the oscillation threshold is inversely proportional to the scaling of the quality factor [12], that in turn is proportional to $N^7$ as discussed in Sec. IV.

In order to demonstrate the advantage of the 6th order EPD over the lower EPD orders, we compare the lasing threshold of the CROW operating near a 6DBE with another CROW of the same length, but different coupling parameters, that is operating near a DBE (4th order EPD) at the same frequency [12], [34]. The threshold in the DBE case is denoted by blue markers in Fig. 5, where the fitting curve of the lasing threshold scaling $\gamma_{th,DBE} \simeq aN^{-5}$ with $N$ is shown with blue dashed line. It is clear that the 6DBE provides lower lasing threshold than the DBE and a stronger scaling with the CROW length, representing a new scaling law for lasing threshold.

### B. Ultra-sensitive optical sensor

Another interesting property of EPDs is that they are very sensitive to perturbations that make them very promising for sensor applications [22], [44]. This also poses strict constraints on fabrication tolerances as already discussed in [13] referring to the 4th order DBE. In order to demonstrate this sensitivity property, we consider the eigenvalues and dispersion relation of the infinite periodic circuit around the 6DBE condition. The tiny physical perturbation of the waveguide (in frequency, refractive index, couplings, losses, dimensions, etc) is modeled by a tiny perturbation $\epsilon$ of the transfer matrix $\underline{\mathbf{T}}$ as $\underline{\mathbf{T}} = \underline{\mathbf{T}}_e + \epsilon \underline{\mathbf{T}}'$, where $\underline{\mathbf{T}}_e$ is the transfer matrix generating the 6DBE and $\underline{\mathbf{T}}'$ accounts for the perturbed terms. In close proximity of the 6DBE, the perturbed eigenvalues of the system characterized by the perturbed transfer matrix $\underline{\mathbf{T}}$ are obtained using the Puiseux series expansion [45], [46] around the ideal degenerate eigenvalue $\zeta_e$, which is approximated to the first order as

$$\zeta_q(\epsilon) \approx \zeta_e + \alpha_1 e^{iq\pi/3}\epsilon^{1/6} \qquad (11)$$

for $q = 1, 2, 3, ..., 6$, where $\zeta_e = \pm 1$ is the 6DBE degenerate eigenvalue, $\epsilon$ is the perturbation factor of any parameter from the designed 6DBE value, and the Puiseux series coefficient $\alpha_1$ is calculated using the formula provided in [46]. It is worth mentioning that $\alpha_1$ depends on both the perturbed system parameter and the flatness of the 6DBE dispersion curve. Since $\zeta_q = \exp(ik_q d)$ and $\zeta_e = \exp(ik_e d)$, the Puiseux series for the eigenvalues $\zeta_q$ leads to the following first order approximation fractional-power expansion of the six Bloch wavenumbers in the vicinity of the 6DBe wavenumber $k_e = 0$ or $k_e = \pi / d$ :

$$k_q(\epsilon) \approx k_e - i\alpha_1 d^{-1} e^{-ik_e d} e^{iq\pi/3}\epsilon^{1/6} \qquad (12)$$

This equation shows that there are 6 different perturbed Bloch wavenumber, and the perturbations $k_q(\epsilon) - k_e$ are proportional to $\epsilon^{1/6}$ which means the wavenumber shows much higher sensitivity for small values of the perturbation $\epsilon$ than the linear proportionality in a regular straight waveguide. Further investigation of the 6DBE sensitivity is beyond the scope of this paper and it can be a subject for future studies including system tolerances and other perturbing effects. We just add that in the case where $\epsilon$ represents a perturbation to the operating 6DBE angular frequency, i.e., $\epsilon = (\omega - \omega_e) / \omega_e$, the relation between $\alpha_1$ and the flatness parameter $\eta_e$ in (5) is $\alpha_1 = k_e d (\eta_e)^{-1/6}$. In the case of $k_e = 0$, $\eta_e$ in (5) would also vanish; hence $\alpha_1$ would assume a finite value. Effect of noise in such a 6DBE system imposes limitations on detection as it has been discussed in [47], [48] in the contest of PT-symmetric systems. The study of theoretical bounds of sensitivity for this class of gainless and (almost) lossless EPD waveguide systems is an important direction for future work.

Finally, we would like to point out that in this paper we only considered lossless structures. Some effect of losses and structural perturbations of the CROW were discussed before in [5], [13]; however such analysis was presented for the DBE case and more studies in this direction shall be performed. In [5], [13] we show that even in the presence of losses and fabrication tolerances, we are still able to obtain in an approximate way the unique properties associated with EPDs, e.g., the scaling of $Q$ factor and of the lasing threshold, if losses are not too large. Losses are expected to have a larger impact in the case of the 6DBE than the DBE case. Furthermore, it is expected that the impact of losses is also affected by the flatness constant $\eta_e$ in (5). This higher sensitivity to losses and other perturbations may be disadvantageous for some applications, but at the same time it definitely represents an outstanding advantage in several applications like optical sensors, modulators and optical switches. In the presence of losses, one may still be able to quantify how close the system is to the 6DBE using the hyperdistance concept explained in Appendix C.

## Conclusion

We used a waveguide coupled to a periodic CROW to show for the first time a possible realization of a 6th order exceptional point of degeneracy at optical frequencies. The EPD studied here belongs to a class of EPDs found in lossless and gainless waveguides. We have investigated the quality factor of a cavity made of a finite-length CROW resonating near the 6DBE frequency and we have shown that the scaling of the loaded $Q$ factor of the 6DBE-CROW cavity is proportional to $N^7$ for lossless CROWs. In addition, we present a significant application of the 6DBE-CROW as a low threshold laser. We have found that the lasing threshold scales with the CROW-cavity length as $g_{th} \propto N^{-7}$ when the CROW operates near the 6DBE, hence it decreases much faster than the threshold of a laser based on a CROW cavity operating near the 4th order DBE, when varying the cavity length. Moreover, we have discussed how the CROW Bloch wavenumbers are perturbed near a 6DBE due to a small structural perturbation $\delta$, and it was shown that the wavenumbers variation is proportional to $\delta^{1/6}$ indicating how a 6DBE-CROW is very sensitive to small perturbations. We expect that the 6DBE in a CROW is affected by perturbations in the system analogously to what was shown in Ref. [13], though further studies would be needed. It is important to point out that even though the EPD is a mathematical critical point, the system retains most of the physical properties of coalescing eigenvectors also when is not operating exactly at the EPD. Indeed, how a system operates close to an EPD can be measured using the "hyperdistance" from an EPD introduced in Ref. [20].

## Appendix A: Lasing Threshold Calculations of the 6DBE-CROW

The lasing threshold of the finite-length 6DBE-CROW forming the cavity is defined as the minimum gain $\gamma = -2k_0 n_{imag}$ [m$^{-1}$] necessary to start and keep oscillations in the CROW cavity. It can be calculated through tracking the poles loci of the 6DBE-CROW transfer function $T_F(\omega)$ in the complex $\omega$-plane defined in (7) and record the gain value at which the system is critically stable, i.e., when a pair of the system poles are exactly located on the real $\omega$-axis. Note that here we mathematically extend the notion of transfer function onto the unstable region to be able to track the poles. An approximate but effective way to track poles crossing the real $\omega$-axis is based on tracking the magnitude of the transfer function $T_F(\omega)$ when varying the distributed gain $\gamma$. The magnitude of $T_F(\omega)$ at an arbitrary complex frequency $\omega$ is inversely proportional to the distance between $\omega$ and the location of the poles $p_i$ in the complex $\omega$-plane as [43]

$$|T_F(\omega)| \propto \frac{1}{\prod_i |(\omega - p_i)|} \propto \frac{1}{\prod_i r_i} \qquad (A1)$$

where $r_i$ is the distances from $\omega$ to the pole $p_i$, as shown in Fig. 6(a). This property arises by using a polynomial expansion of the transfer function denominator in the neighborhood of $\omega = \omega_{r,e}$. Note that since the electric field is a real-valued quantity, poles occur in pairs with the same imaginary part while the real parts have opposite signs, i.e., if $p_i$ is a pole of the system, then $-p_i^*$ is also a pole, where the star means complex conjugation. In Fig. 6(a) we show only the right half of the complex $\omega$-plane for simplicity.

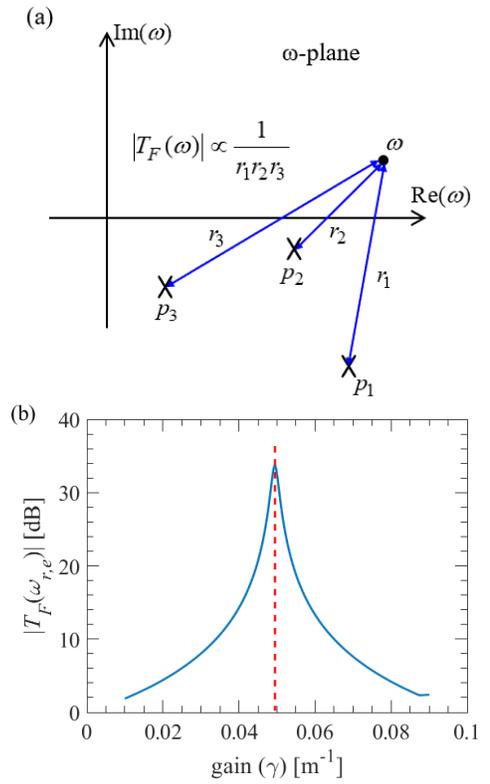

Fig. 6. (a) Example of three poles in the complex $\omega$-plane, of the transfer function $T_F$. The magnitude of $T_F$ at an arbitrary frequency $\omega$ is inversely proportional to the product of "distances" between $\omega$ and the poles. (b) Plot of of the magnitude of the transfer function, $|T_F(\omega)|$, defined in Eq. (7), of a 10 unit cell finite-length periodic 6DBE-CROW calculated at the 6DBE resonance $\omega = \omega_{r,e}$ versus distributed gain introduced in the rings of the 6DBE-CROW cavity. The oscillation (lasing) threshold is approximately the gain value at which the $|T_F(\omega_{r,e})|$ is maximum and shown with the red dashed line.

In the absence of gain, the system is stable and since we are adopting the time convention as $e^{-i\omega t}$, poles associated to a stable system must have negative imaginary part, i.e., the poles are in the lower half of the $\omega$-plane. Adding gain to the system results in a migration of the poles towards the upper half of $\omega$-plane with positive imaginary part.

We choose to calculate $\left| T_F(\omega) \right|$ at the 6DBE resonance frequency, i.e., at $\omega = \omega_{r,e}$ that is a purely real frequency by definition (corresponding to the peak of $T_F(\omega)$ in Fig. 3), and represented by a point on the real $\omega$-axis. If the system is passive (with some internal losses and with termination losses) poles are such that $\text{Im}(p_i) < 0$, and when we add gain eventually a pair of poles will migrate to the upper half of the complex $\omega$-plane such that $\text{Im}(p_i) > 0$. When $\text{Im}(p_i) = 0$ for a pair of poles that cross the real axis, when increasing the gain, the "distances" of these poles from the real $\omega_{r,e}$ experience a local minimum and the magnitude of the transfer function, $\left| T_F(\omega_{r,e}) \right|$, exhibits a local maximum. This is an approximate but simple method to figure out when a pole crosses the real $\omega$ axis. Under the condition $\text{Im}(p_i) = 0$ the system is said critically stable, and the gain that renders $\text{Im}(p_i) = 0$ is the threshold gain.

In Fig. 6 (b) we show the magnitude of the transfer function of a finite-length 6DBE-CROW made of 10 unit cells defined as in Eq. (7) and calculated at the 6DBE-resonance frequency, i.e., $T_F(\omega = \omega_{r,e})$, versus gain. The $\left| T_F(\omega_{r,e}) \right|$ is maximum at a specific gain value which, based on what we have just discussed, is the threshold gain, shown in Fig. 6(b) by the red dashed line. Note that the concept of transfer function is properly defined only for stable systems, since above threshold a system is unstable and it would start to oscillate. Here we plot $\left| T_F(\omega_{r,e}) \right|$ also for gain values above the threshold just to understand when the minimum distance occurrs. Also, note that the $\left| T_F(\omega_{r,e}) \right|$ will show another local maxima when we further increased gain, corresponding to other poles crossing the real $\omega$-axis, however we only care about the first local maximum because the system would start oscillating when the first pole crosses the real $\omega$-axis.

## Appendix B: Transfer Matrix of a 6DBE-CROW Unit Cell

In order to write an expression for the transfer matrix of the 6DBE-CROW unit cell shown in Fig. 1(b) of length $d = 4R$, we divide the unit cell into two segments 1 and 2 (they are identical except for the coupling coefficients) each of length $2R$ as shown in Fig. 7. Note that the state vectors $\boldsymbol{\psi}(z)$, $\boldsymbol{\psi}(z + d/2)$, and $\boldsymbol{\psi}(z + d)$ are defined at the immediate left of each truncation, just before where the ideal couplings $\kappa_1$ and $\kappa_2$ occur. The T-matrix of each segment is obtained by first writing the scattering matrix (S-matrix) of each segment assuming single mode propagation in each segment (for each direction) as done in Appendix A in [5], and then by transforming the obtained S-matrix into the desired T-matrix of the segment. The T-matrix of segment 1 (the left half of the unit cell) is then given by

$$\underline{\underline{T}}_1(z, z + 2R) = \begin{pmatrix} \underline{\underline{T}}_{11} & \underline{\underline{T}}_{12} \\ \underline{\underline{T}}_{21} & \underline{\underline{T}}_{22} \end{pmatrix} \quad (B1)$$

with

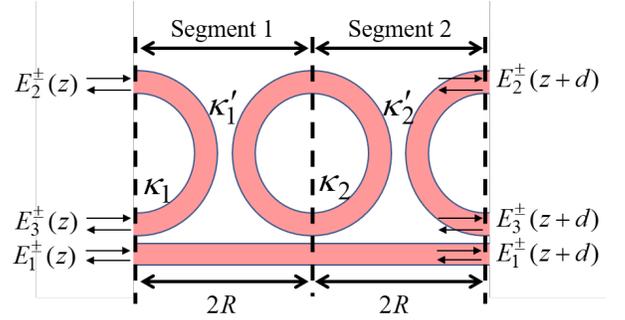

Fig. 7. A unit cell of the periodic 6DBE-CROW of length $d = 4R$ with the electric field wave amplitudes defined at the periodic-cell boundaries. The unit cell is consisting of two identical segments, each of length $2R$, except for the coupling coefficients; in segment 1 the coupling coefficients are $\kappa_1$ and $\kappa'_1$, while in segment 2 they are $\kappa_2$ and $\kappa'_2$. The dashed lines marking the start and end of each segment coincides with the vertical symmetry line of the corresponding ring resonator and they are defined just before the coupling point between the ring resonator and the straight waveguide.

$$\underline{\underline{T}}_{11} = \underline{\underline{T}}_{22}^* = \begin{pmatrix} \sqrt{1 - \kappa_1^2}\, e^{i2n_w k_0 R} & \dfrac{-\kappa_1}{\kappa'_1} e^{i n_r k_0 z} & 0 \\ 0 & 0 & \dfrac{i e^{i\pi n_r k_0 R}}{\kappa'_1} \\ i\kappa_1 e^{i2n_w k_0 R} & \dfrac{i\sqrt{1 - \kappa_1^2}}{\kappa'_1} e^{i\pi n_r k_0 R} & 0 \end{pmatrix} \quad (B2)$$

$$\underline{\underline{T}}_{12} = \underline{\underline{T}}_{21}^* = \dfrac{\sqrt{1 - \kappa_1'^2}}{\kappa'_1} \begin{pmatrix} 0 & 0 & \kappa_1 \\ 0 & -i & 0 \\ 0 & 0 & -i\sqrt{1 - \kappa_1^2} \end{pmatrix}, \quad (B3)$$

where all the parameters are defined in Sec. II and the * symbol denotes the conjugate operation. One can easily obtain the T-matrix of segment 2 (right half of the unit cell shown in Fig. 7) through replacing the coupling coefficients $\kappa_1, \kappa'_1$ with $\kappa_2, \kappa'_2$. Then we find the unit cell matrix $\underline{\underline{T}}$ by multiplying the T-matrices of the two segments: $\underline{\underline{T}} = \underline{\underline{T}}_2 \underline{\underline{T}}_1$.

We want to stress that in the absence of loss and gain, the T-matrix $\underline{\underline{T}}$ describing mode propagation across the unit cell is not Hermitian. Furthermore, $\underline{\underline{T}}$ of the lossless CROW follows the fundamental J-unitary property [8], [20]; which means that $\underline{\underline{T}}^{-1}(z_2, z_1) = \underline{\underline{J}}\,\underline{\underline{T}}^\dagger(z_2, z_1)\,\underline{\underline{J}}^{-1}$ with the matrix $\underline{\underline{J}}$ defined as a matrix that satisfies $\underline{\underline{J}} = \underline{\underline{J}}^{-1} = \underline{\underline{J}}^\dagger$ where the dagger $\dagger$ denotes complex conjugation and transpose operation [49]. An example of the matrix $\underline{\mathbf{J}}$ for the CROW in this paper is given by

$$\underline{\mathbf{J}} = \begin{pmatrix} 1 & 0 & 0 & 0 & 0 & 0 \\ 0 & 1 & 0 & 0 & 0 & 0 \\ 0 & 0 & 1 & 0 & 0 & 0 \\ 0 & 0 & 0 & -1 & 0 & 0 \\ 0 & 0 & 0 & 0 & -1 & 0 \\ 0 & 0 & 0 & 0 & 0 & -1 \end{pmatrix} \quad \text{(B4)}$$

The sixth order EPD is induced in the CROW due to the evanescent coupling between the waveguides and the periodicity of the structure so that at the EPD frequency $\underline{\mathbf{T}}$ is similar to a Jordan block matrix of order six.

### Appendix C: Eigenvector Coalescence Using the Hyperdistance Calculations Near a 6th Order EPD

The hyperdistance concept described in [20] is used here to show how a system is approaching an EPD based on the coalescence of eigenvectors. As a system is subject to various types of perturbations, e.g., losses, fabrication tolerances or frequency detuning, the system may deviate away from the EPD. Hence, the hyperdistance denoted by $D_H$ is a measure of the coalescence of the system eigenvectors and can be defined as

$$D_H = \frac{1}{15} \sum_{\substack{m=1,n=2 \\ n>m}}^{6} \left| \sin\left(\theta_{mn}\right) \right|, \quad \cos\left(\theta_{mn}\right) = \frac{\left| \langle \psi_m, \psi_n \rangle \right|}{\|\psi_m\| \|\psi_n\|} \quad \text{(C1)}$$

where $\theta_{mn}$ is the angle between the two $m$th and $n$th eigenvectors in a six-dimensional complex vector space. Such angle is defined via the magnitude of the inner product $\left| \langle \psi_m, \psi_n \rangle \right| = \psi_m^\dagger \psi_n$, where $\|\psi_m\|$ and $\|\psi_n\|$ are the norms of the complex valued eigenvectors $\psi_m$ and $\psi_n$, respectively. The hyperdistance $D_H$ takes a value

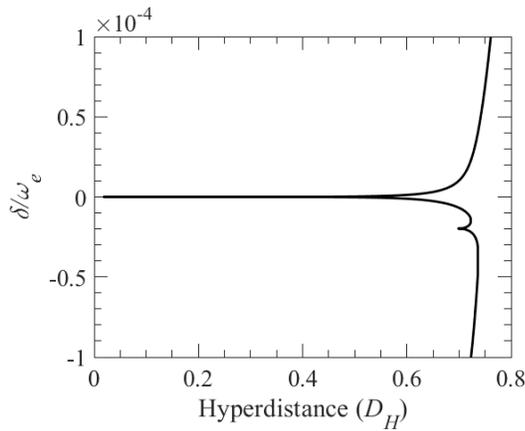

Fig. 8. The hyperdistance $D_H$ of the periodic 6DBE-CROW vanishing at the 6DBE designed angular frequency, corresponding to the wavelength $\lambda_e = 1550$ nm. Here $\delta = \omega - \omega_e$. When $\omega = \omega_e$, one has $D_H \approx 0$ which indicates the coalescence of all the six eigenvectors at the 6DBE frequency.

between 0 and 1, and for the ideal coalescence of six eigenvectors, i.e., exactly at the 6DBE, one has $D_H = 0$. In the case where structural perturbations are present, $D_H \neq 0$ and one may define a threshold value of the hyperdistance below which the system exhibits the unique characteristics associated with the existence of EPDs.

In Fig. 8 we show the calculated hyperdistance for the CROW dispersion diagram shown in Fig. 2. We clearly see that when approaching the designed EPD frequency, i.e., when $\delta / \omega_e \triangleq (\omega - \omega_e) / \omega_e \to 0$, one has $D_H \to 0$, that confirms the coalescence of all the six eigenvectors at that frequency. What shown here could be used also to discuss how close the waveguide system is to an EPD when losses and fabrication tolerances are present.

### Acknowledgment

This material is based on work supported by the Air Force Office of Scientific Research award number FA9550-15-1-0280 and by the National Science Foundation under award NSF ECCS-171197.